\RequirePackage{ifpdf}
\ifpdf 
\documentclass[pdftex]{sigma}
\else
\documentclass{sigma}
\fi

\def\ba{\begin{array}{c}}
\def\ea{\end{array}}

\newcommand{\R}{\mathbb{R}}

\newcommand{\be}{\begin{equation}}
\newcommand{\ee}{\end{equation}}
\newcommand{\ben}{\begin{displaymath}}
\newcommand{\een}{\end{displaymath}}
\newcommand{\bea}{\begin{eqnarray}}
\newcommand{\eea}{\end{eqnarray}}
\newcommand{\nn}{\nonumber}
\newcommand{\kt}{\rangle}
\newcommand{\br}{\langle}
\newcommand{\bbr}{\br\!\br}

\newcommand{\pbr}{\prec\!}
\newcommand{\pkt}{\!\succ}

\begin{document}

\renewcommand{\thefootnote}{$\star$}

\renewcommand{\PaperNumber}{085}

\FirstPageHeading

\ShortArticleName{Cryptohermitian Picture of Scattering Using Quasilocal
Metric Operators}

\ArticleName{Cryptohermitian Picture of Scattering\\ Using Quasilocal
Metric Operators\footnote{This paper is a
contribution to the Proceedings of the 5-th Microconference
``Analytic and Algebraic Me\-thods~V''. The full collection is
available at
\href{http://www.emis.de/journals/SIGMA/Prague2009.html}{http://www.emis.de/journals/SIGMA/Prague2009.html}}}

\Author{Miloslav ZNOJIL}

\AuthorNameForHeading{M. Znojil}

\Address{Nuclear Physics Institute ASCR,
250 68 \v{R}e\v{z}, Czech Republic}

\Email{\href{mailto:znojil@ujf.cas.cz}{znojil@ujf.cas.cz}}
\URLaddress{\url{http://gemma.ujf.cas.cz/~znojil/}}

\ArticleDates{Received July 05, 2009, in f\/inal form August 23, 2009;  Published online August 27, 2009}

\Abstract{One-dimensional unitary scattering controlled by
non-Hermitian (typically, ${\cal PT}$-symmetric) quantum
Hamiltonians  $H\neq H^\dagger$ is considered. Treating these
operators via Runge--Kutta approximation, our three-Hilbert-space
formulation of quantum theory is reviewed as explaining the
unitarity of scattering. Our recent paper on bound states [\mbox{\href{http://dx.doi.org/10.3842/SIGMA.2009.001}{\mbox{Znojil~M.}, {\it SIGMA} {\bf 5} (2009), 001, 19~pages}}, \href{http://arxiv.org/abs/0901.0700}{arXiv:0901.0700}] is complemented by the
text on scatte\-ring.  An elementary example illustrates the
feasibility of the resulting innovative theoretical recipe. A new
family of the so called quasilocal inner products in Hilbert space
is found to exist. Constructively, these products are all described
in terms of certain non-equivalent short-range metric operators
$\Theta\neq I$ represented,  in Runge--Kutta approximation, by
$(2R-1)$-diagonal matrices.}


\Classification{81U20; 46C15; 81Q10; 34L25; 47A40; 47B50}

\section{Introduction and summary}

In paper I \cite{SIGMA} we summarized the situation in which certain
very complicated quantum bound-state Hamiltonians
$\mathfrak{h}=\mathfrak{h}^{\rm (P)}=\mathfrak{h}^\dagger$ (acting in
the usual physical Hilbert space ${\cal H}^{\rm (P)}$) were assumed
solvable via a Dyson-inspired transition to their non-Hermitian but
perceivably simpler isospectral partners
$H=\Omega^{-1}\mathfrak{h}\Omega \neq H^\dagger$. The latter
auxiliary operators were assumed acting in a~doublet of the f\/irst
auxiliary and the second auxiliary Hilbert spaces ${\cal
H}^{\rm (F,S)}$, respectively. The latter space ${\cal H}^{\rm (S)}$ was
f\/inally assumed endowed with an unusual inner product $(a,b)^{\rm (S)} $
def\/ined by the formula $(a,b)^{\rm (S)} := (a,\Theta   b)^{\rm (F)}$ in
terms of the usual Dirac's inner product $(a,b)^{\rm (F)} $ and of the
so called metric operator
$\Theta=\Theta^\dagger=\Omega^\dagger\Omega>0$.

In the present continuation of paper I we shall extend our
three-Hilbert-space formulation of quantum theory to a class of
models of scattering. For illustrative purposes we shall discretize
the axis of coordinates into Runge--Kutta lattice \cite{RK}. This
discretization is not a mandatory ingredient of the approach but its
use will facilitate our explicit constructions. In particular, it
will enable us to demonstrate that for a selected sample Hamiltonian
$H$ one can construct several alternative metrics $\Theta$ which are
all compatible with the unitarity of the scattering-state solutions
of the underlying Schr\"{o}diner equation. This quantif\/ication of
the ambiguity of $\Theta$ will be a core of our message to
physicists.
Unexpectedly, our unitarity-supporting illustrative $\Theta$s will
prove ``quasilocal", i.e., they will emerge as short-ranged
continuous-coordinate limits of $(2R-1)$-diagonal Runge--Kutta
approxi\-mation matrices $\Theta=\Theta_R$ with integer $R=R(h)$ such
that $\lim\limits_{h\to 0}hR(h) =0$. This will be our main new mathematical
result.

The detailed presentation of this result will be preceded by a short
review of motivation (Section~\ref{context}) and of the currently
well-established theoretical understanding of non-Hermitian models
of bound states (Section~\ref{horizons}). The current status of
extension of this version of quantum theory to scattering scenario
will be summarized in Section~\ref{problems}.

In Section~\ref{core} we pick up one of the elementary concrete
models~\cite{FT} and use it to outline our method applicable in the
dynamical regime of scattering. The core of our message is
formulated in this section. We f\/ix there the level of approximation
given by the Runge--Kutta lattice-spacing $h>0$ and describe the
construction of metric operators $\Theta=\Theta(H)$. For our toy
model $H$, in particular, we list the f\/irst few explicit samples of
$\Theta=\Theta_R(H)$ possessing the $(2R-1)$-diagonal matrix
structure. These matrices are constructed by computerized symbolic
manipulations at $R = 1,2,\ldots,7$. Our knowledge of these
solutions enables us to conjecture (and, subsequently, prove)
formula which def\/ines all the eligible quasilocal metrics
$\Theta_R(H)$ at any $R$ in closed form.

In complementary and concluding Sections~\ref{dyson} and
\ref{summary} we construct underlying Dyson opera\-tors~$\Omega$ and
discuss some of their potentially most relevant descriptive
properties as well as several possible consequences of their use in
model-building.

\section[Context: ${\cal PT}$-symmetric Hamiltonians]{Context: $\boldsymbol{\cal PT}$-symmetric Hamiltonians}\label{context}

Among all of the simplif\/ied phenomenological models used in quantum
mechanics a prominent role is played by the families described by
Schr\"{o}dinger equations of ordinary dif\/ferential form
 \be
 \frac{  \hbar^2}{2m}
  \left [-\frac{d^2}{d{x}^2}
  +  \frac{L(L+1)}{{x}^2}\right ]  \Psi({x})
 +V(x)   \Psi({x}) = E  \Psi({x}) .
 \label{SEor}
  \ee
Centrifugal coef\/f\/icient $L(L+1)$ may be admitted to vanish. Variable $x$ is then interpreted as the one-dimensional coordinate.
Requirement $\Psi(x) \in {\mathbb L}_2({\mathbb R})$ is being imposed upon
bound states while the scattering solutions are characterized by
asymmetric asymptotic boundary conditions at all real $\kappa =
\sqrt{E-V(\infty)}$,
 \be
 \Psi(x) =
 \left \{
 \begin{array}{ll}
 e^{i\kappa x}+R e^{-i\kappa x} ,\quad
 & x \ll -1 ,\\
 T e^{i\kappa x} ,\quad
 &x \gg 1 .
 \ea
 \right .
 \label{scatbc}
 \ee
At nonvanishing centrifugal coef\/f\/icients one demands that
$L(L+1)>-1/4$ while variable $x$ is treated as the radial coordinate
in $D$ dimensions, $x= |\vec{r}| \in (0,\infty)$. An additional,
specif\/ic boundary condition must be then imposed in the origin in
order to guarantee the regularity of  bound-state as well as
scattering solutions \cite{Znopra}. Of course, physics represented
by equation~(\ref{SEor}) on half line depends again on the asymptotic
behavior of potential $V(x)$.

Usually one distinguishes between the conf\/ining, bound-state regime
(where $V(x)$ is chosen smooth, real and very large at large $x \to
\pm \infty$) and the scattering scenario (where, typically, $V(\pm
\infty)=0$ and where wave functions $\Psi({x})$ are selected as free
waves at large $x \to \pm \infty$). Recently, several authors
emphasized the existence of another, unusual quantization of the
ordinary dif\/ferential Schr\"{o}dinger equation~(\ref{SEor}) which is
assumed integrated along a non-standard, {\em complex} contour of
``pseudocoordinates" $x=x(s)\in {\mathbb C}$ where $s \in {\mathbb R}$ is a
parameter (cf., e.g., Sibuya \cite{Sibuya} or Bender and Turbiner
\cite{BT}).

The early attempts in this direction were treated and accepted as a
mere mathematical curiosity. For illustration we may recall paper
\cite{BG} by Buslaev and Grecchi who studied anomalous potentials
$V(x) \sim -x^4 + {\cal O}(x^3)$. Equation (\ref{SEor}) with these
``wrong-sign" potentials has been considered along a shifted line of
$x=x^{\rm (BG)}(s)=s-{\rm i}\varepsilon$. By this trick, the potential
(which looks asymptotically repulsive) has been made manifestly
conf\/ining. In parallel, the usual requirement of Hermiticity has
been replaced by ${\cal PT}$-symmetry, i.e., by left-right symmetry
in the complex plane of $x$ (cf.\ also~\cite{Mateo} for more
comments).

During the last ten years we witnessed a quick growth of interest in
${\cal PT}$-symmetric models. Within physics community this interest
has been inspired by Bender's and Boettcher's inf\/luential letter
\cite{BB} where the authors emphasized the deeply physical appeal of
the combination of the parity-reversal symmetry (mediated by the
operator ${\cal P}$) with the time-reversal symmetry (cf.\ operator
${\cal T}$). They conjectured and demonstrated, by approximate
methods, the reality (i.e., in principle, measurability) of the
bound-state spectra for the whole family of specif\/ic ${\cal PT}$-symmetric toy potentials
 \be
  V^{\rm (BB)}(x) = x^{2} ({\rm i}x)^{4\delta} , \qquad \delta
  \geq 0 .
  \label{BBmod}
  \ee
A few years later the validity of the conjecture has rigorously been
proved by Dorey, Dunning and Tateo~\cite{DDT} and by Shin~\cite{Shin}. This clarif\/ied, for bound states, the mathematics of
complexif\/ication of the argument $x$ of  wave functions $\psi(x)$
and/or of potential~$V(x)$.

The questions emerging in scattering regime have temporarily been
left open.

\section{New horizons: cryptohermitian observables}\label{horizons}

The success of the latter generalization of the class of quantum
bound-state models opened several new theoretical directions of
research since in many concrete examples the complex value of
``coordinate'' $x$ ceased to be tractable as a measured or measurable
quantity. For this reason the theoretical as well as practical
acceptance of the internal mathematical consistency of apparently
non-Hermitian models exemplif\/ied by equations~(\ref{SEor}) +
(\ref{BBmod}) took some time in physics community (cf., e.g., review
paper~\cite{Carl} for historical comments). The paradoxes seem to be
clarif\/ied at present and, for bound states, a new pattern is
established for implementation of the formalism of textbook quantum
mechanics.

The key to the safe return from equations~(\ref{SEor}) + (\ref{BBmod})
(and the like) to textbooks can be seen in the concept of
cryptohermiticity (the word invented, very recently, by Smilga
\cite{Smilga}) which, in essence, means that an operator $H$ which
appears non-Hermitian in Hilbert space ${\cal H}^{\rm (F)}$ (where the
superscript indicates the (user-)friendliness of the most current
mathematical def\/inition of the inner product -- see below) may be
reinterpreted, under certain circumstances, as safely Hermitian in
another, Hilbert space ${\cal H}^{\rm (S)}$ (where the superscript
stands for ``standard'' physics). The third Hilbert space ${\cal
H}^{\rm (P)}$ emerges, quite naturally, as a space which is unitarily
equivalent to~${\cal H}^{\rm (S)}$ (i.e., it represents {\em strictly
the same} physics so that for the purposes of physical predictions
we have an entirely free choice between ${\cal H}^{\rm (P)}$ and  ${\cal
H}^{\rm (P)}$) but which is {\em the only} space encountered in
textbooks (i.e., the metric operator in ${\cal H}^{\rm (P)}$ remains
conventional, identically equal to the unit operator).

One of the f\/irst papers presenting and reviewing such an abstract
idea in a concrete application in nuclear physics appeared more than
f\/ifteen years ago~\cite{Geyer}. The study of heavier atomic nuclei
has been shown simplif\/ied by the mapping of the usual, complicated
physical Hilbert space of states ${\cal H}^{\rm (P)}$ (describing
fermions and possessing, therefore, complicated antisymmetrization
features) upon another, bosonic (i.e., manifestly {\em unphysical})
auxiliary space ${\cal H}^{\rm (F)}$. This made the calculations
perceivably facilitated. Whenever needed, the return to fermionic
wave functions mediated by the not too complicated Dyson mapping
$\Omega$ remained feasible via a unitary equivalence between ${\cal
H}^{\rm (P)}$ and ${\cal H}^{\rm (S)}$ (more details may be found
in~\cite{SIGMA}).

The combination of a comparatively narrow applicability in nuclear
physics with a rather unusual mathematics caused that the methodical
appeal of the non-unitary-mapping idea remained virtually unnoticed
until its re-emergence in ${\cal PT}$-symmetric context. At present
we witness a~massive revival of interest in the parallel use of the
three alternative representations~${\cal H}^{\rm (P)}$, ${\cal H}^{\rm (F)}$
and ${\cal H}^{\rm (S)}$ of the {\em same} physical quantum system. The
invertible Dyson-type mapping~$\Omega$ relates the spaces ${\cal
H}^{\rm (P)}$ and ${\cal H}^{\rm (F)}$ as well as the respective
Hamiltonians or operators of other observables.

The manifest violation of the unitarity by the mapping $\Omega\neq
(\Omega^\dagger)^{-1}$ requires due care in applications. For
example, the physical Hamiltonian $\mathfrak{h}$ must be Hermitian
(i.e., more strictly, essentially self-adjoint in the physical
Hilbert space ${\cal H}^{\rm (P)}$ exemplif\/ied by the fermionic space in
the above-mentioned concrete application). It may also be perceived
as a transform of another operator def\/ined as acting in another
vector space, $\mathfrak{h} = \Omega H \Omega^{-1}$. The latter
``Hamiltonian'' $H$ is, by construction, def\/ined {\em and} manifestly
non-Hermitian in  space ${\cal H}^{\rm (F)}$ where, trivially,
 \be
 H^\dagger =\Omega^\dagger \mathfrak{h} \big(\Omega^{-1}\big)^\dagger
 =\Omega^\dagger \Omega H \Omega^{-1}\big(\Omega^{-1}\big)^\dagger
 = \Theta H \Theta^{-1} .
 \label{crypto}
 \ee
We abbreviated $\Theta =\Omega^\dagger \Omega$ calling this new
operator, conveniently, a metric. It is def\/ined as acting on the
kets in both Hilbert spaces ${\cal H}^{\rm (F)}$ and ${\cal H}^{\rm (S)}$.
This mathematical ambiguity proves inessential in applications and
it is also easily clarif\/ied by the following elementary graphical pattern using notation of paper~I,
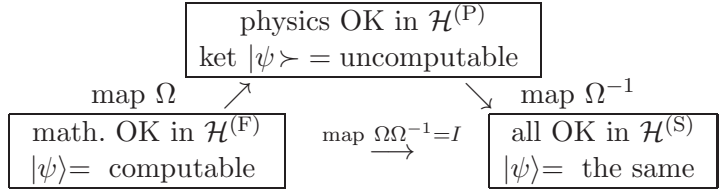
\[
  \ba
    \begin{array}{|c|}
 \hline
  \ {\rm physics\ OK\  in} \
 {\cal H}^{\rm (P)}\ \\
    {\rm ket}\ |\psi\pkt
  {\rm  = uncomputable} \ \\
 \hline
 \ea
\\ {\rm map}\
 \Omega \ \ \ \   \nearrow \ \  \  \ \ \ \ \ \
 \ \ \ \ \ \ \ \
\ \ \ \ \   \searrow  \ \ {\rm map}\  \Omega^{-1}\\
 \begin{array}{|c|}
 \hline
 \ {\rm math.\ OK\ in\ }
 {\cal H}^{\rm (F)}\ \\
   |\psi\kt
  {\rm =\ computable}\  \\
  \hline
 \ea\ \ \
 \stackrel{{\rm map}\ \Omega  \Omega^{-1}=I}{ \longrightarrow }
 \ \
 \begin{array}{|c|}
 \hline
 {\rm all \ OK\ in}\
 {\cal H}^{\rm (S)} \\
   |\psi\kt
  {\rm =\ the \ same} \ \\
 \hline
 \ea
\\
\ea
\]
This diagram reconf\/irms that the Hermiticity of the physical
Hamiltonian $\mathfrak{h} =\Omega H \Omega^{-1}
=\mathfrak{h}^\dagger= \left [\Omega^{-1}\right
]^\dagger H^\dagger \Omega^\dagger$ acting in the physical Hilbert
space ${\cal H}^{\rm (P)}$ is equivalent to the cryptohermiticity
constraint (\ref{crypto}) imposed upon its partner
$H=\Omega^{-1} \mathfrak{h} \Omega \neq H^\dagger$ in ${\cal
H}^{\rm (F)}$.

\subsection[The ambiguity of the metric $\Theta=\Theta(H)$]{The ambiguity of the metric $\boldsymbol{\Theta=\Theta(H)}$}\label{Ib}

Hesitations may emerge when one imagines that the Hermitian
conjugation itself is not a unique operation and that it may be
altered \cite{Messiah}. In this way equation~(\ref{crypto}) may be
interpreted as assigning several alternative physical metric
operators to single Hamiltonian, $\Theta\in
(\Theta_1(H),\Theta_2(H), \ldots)$. The f\/irst thorough discussion
and clarif\/ication of this problem of ambiguity has been published by
Scholtz et al.~\cite{Geyer}. They emphasized that besides the
Hamiltonian $H$ itself, {any other} operator ${\cal O}={\cal
O}_1,{\cal O}={\cal O}_2,\ldots$ of an observable quantity must obey
the {same} cryptohermiticity relation as $H\ \equiv\ {\cal O}_0$
itself. In opposite direction, for a given set of observables in
${\cal H}^{\rm (F)}$, each eligible metric operator $\Theta$ must remain
compatible with all of them,
 \be
   \Theta\,{\cal O}_j =  {\cal O}^\dagger_j \Theta ,\qquad
 j=(0,)  1, 2, \ldots,j_{\max}.
 \label{newhtot}
 \ee
These requirements reduce the ambiguity of $\Theta$ so that at a
suitable integer $j_{\max}$ the metric $\Theta=\Theta(H)$ may become,
in principle, unique.

The ambiguity of $\Theta$ may acquire an enormous theoretical
importance as well as phenomenological relevance.  The abstract
understanding of the ambiguity of Dyson maps $\Omega$ did already
prove crucial not only for the above-mentioned  ef\/f\/icient
description of bound states in nuclear physics  but also in a
clarif\/ication of the role of the charge or ghosts in f\/ield theory
\cite{BBJ}, etc.

\subsection{Classical limit in the case of cryptohermitian Hamiltonians }

Persuasive demonstration of relevance of the apparently purely
mathematical subtlety of the ambiguity of metric can be mediated by
the popular Bessis' and Zinn-Justin's (BZJ, \cite{DB}) imaginary
cubic oscillator
 \be
 H^{\rm (BZJ)} |\psi_n^{\rm (F)}\kt =  E_n |\psi_n^{\rm (F)}\kt ,\qquad
 H^{\rm (BZJ)} = -\frac{d^2}{dx^2} + {\rm i} g x^3 ,\qquad  n = 0, 1, \ldots
 \label{daniel}
 \ee
for which the quantum bound-state problem is formulated and solved
in  Hilbert space ${\cal H}^{\rm (F)} \equiv L^2(\R)$ {\em and} for
which the spectrum is real, discrete and bounded below despite the
BZJ Hamiltonian being manifestly non-Hermitian.

Routinely, the Dyson mapping  may be employed to convert the
elements $|\psi^{\rm (F)}\kt \in {\cal H}^{\rm (F)}$ of the above-mentioned
space into their images $ |\psi^{\rm (P)}\kt \in {\cal H}^{\rm (P)}$ which
lie in the correct representa\-tion~${\cal H}^{\rm (P)}$ of the abstract,
textbook Hilbert space of states. Superscript $^{\rm (P)}$ stands for
``physical'' and we have $\Omega:{\cal H}^{\rm (F)} \to {\cal H}^{\rm (P)}$
and $ |\psi^{\rm (P)}\kt= \Omega |\psi^{\rm (F)}\kt$. It is the isospectral
partner $\mathfrak{h}^{\rm (BZJ)}=\Omega H^{\rm (BZJ)} \Omega^{-1}$ of our
original Hamiltonian which becomes, by construction, safely
Hermitian in physical space ${\cal H}^{\rm (P)}$.

There is a price to be paid for the clarif\/ication of theoretical
concepts. In our illustrative example~(\ref{daniel}) it was
dif\/f\/icult to prove that the spectrum of the bound-state energies is
all real~\cite{DDT}. In~\cite{cubic} one f\/inds that even in the
weak-coupling regime with a very small value of $g=\epsilon$ one
arrives at an impressively complicated physical representative of
Hamiltonian in ${\cal H}^{\rm (P)}$,
\begin{gather}
    \mathfrak{h}^{\rm (BZJ)}
     = \Omega H^{\rm (BZJ)} \Omega^{-1}=\frac{p^2}{2}+\frac{3}{16}
    \left(\left\{x^6,\frac{1}{p^2}\right\}
    +22\left\{x^4,\frac{1}{p^4}\right\}+(510+10\lambda_1)
    \left\{x^2,\frac{1}{p^6}\right\}\right.\nn\\
\left.\phantom{\mathfrak{h}^{\rm (BZJ)}=}{} +
    \frac{8820+140\lambda_1}{p^8}
    -\frac{4}{3} \kappa_1\left\{x^3,\frac{1}{p^5}\right\} {\cal P}\right)
    \epsilon^2\nn\\
\phantom{\mathfrak{h}^{\rm (BZJ)}=}{} +\frac{1}{4}\left(
    15\lambda_2 \left(\left\{x^2,\frac{1}{p^{11}}\right\}
    +\frac{44}{p^{13}} \right)-i\kappa_2\left\{x^3,\frac{1}{p^{10}}\right\}
    {\cal P}\right)\epsilon^3+
    {\cal O}(\epsilon^4) .
    \label{he-man}
\end{gather}
This formula contains parity operator ${\cal P}$ and anticommutators
$\{\cdot,\cdot\}$, it uses the Fourier-trans\-for\-med
multiplication-operator representation of the powers of $p$ in
$L^2(\R)$ and, f\/inally, it varies freely with four real parameters
$\lambda_j$ and $\kappa_j$, $j=1,2$. Persuasively, formula
(\ref{he-man}) demonstrates that the {\em explicit} form of true
Hamiltonians  $\mathfrak{h}=\mathfrak{h}^\dagger$ may become
threateningly or even prohibitively complicated. One cannot be
expected to perform practical calculations using this representation
of the Hamiltonian.

The latter remark seems to imply that the explicit knowledge of maps
$H \leftrightarrow \mathfrak{h}$ or, for other observables, $A
\leftrightarrow \mathfrak{a}$ may be skipped as redundant. Actually,
this is just a partial truth since operators $\mathfrak{h}$ or
$\Omega$ may remain unavoidable, e.g., during the analysis of
time-dependent systems (cf.~\cite{timedep} and also the recent
preprint~\cite{Bila} in this respect). Moreover, in model
(\ref{he-man}) one could select $\lambda_2=\kappa_2=0$ in order to
preserve ${\cal PT}$-symmetry. Last but not least, the ``unfriendly"
Hilbert space ${\cal H}^{\rm (P)}$ opens the way towards the classical
limit of the system. From the imaginary cubic example (\ref{he-man})
one arrives at the classical Hamiltonian function $h_c$ def\/ined in
the classical single-particle phase space of the coordinate $p_c$
and momentum $p_c$,
    \be
    h_c^{\rm (BZJ)}=\frac{p^2_c}{2m}+\epsilon^2 g(p_c) {x_c^6}
    +{\cal O}(\epsilon^4) ,\qquad
     g(p_c)=
    \frac{3 m}{8 p_c^2}
    \label{H-classical}
    \ee
(cf.\ equation~(62) in \cite{cubic}). Thus, the limiting transition $\hbar
\to 0$  reveals a hidden sextic-oscillator nature of the imaginary
cubic forces in the weak-coupling regime.

\subsection{Extended Dirac's bra/ket notation}

The popular emphasis on the non-Hermiticity  of {simplif\/ied,
artif\/icial} $H\neq H^\dagger$ in ${\cal H}^{\rm (F)}$ should be
perceived as attracting attention but slightly misleading. It
ref\/lects just the mathematical property (implicitly, of the only
relevant, correct and physical operator $\mathfrak{h}$) with no
direct connection, say, to the principle of correspondence (cf.\ the
preceding paragraph). In contrast to the descriptions of open
systems \cite{Rotter} where the probability is not conserved the
calculations of probabilities must {\em all} be performed inside
${\cal H}^{\rm (P)}$ or in one of its unitarily equivalent partner
Hilbert spaces ${\cal H}^{\rm (S)}$ where $\Theta \neq I$. We strongly
recommend an explicit reference to the space in which one resides,
because
\begin{itemize}\itemsep=0pt

 \item
the apparent and misleading conf\/lict between the absence of the
correct physics in ${\cal H}^{\rm (F)}$ and the lack of the
computational feasibility in ${\cal H}^{\rm (P)}$ is naturally resolved
by the metric-dependent description of the system in the
standardized Hilbert space ${\cal H}^{\rm (S)}$;

 \item
the Dirac notation can be used after a graphical adaptation of the
bra- and ket-symbols when one speaks about the elements of the
formally dif\/ferent Hilbert spaces ${\cal H}^{\rm (F)}$, ${\cal H}^{\rm (S)}$
and ${\cal H}^{\rm (P)}$.

\end{itemize}

More explicitly, the {\em same} state $\psi$ of a quantum system
will be characterized by a spiked ket symbol in $P$-space,
$|\psi\pkt \in {\cal H}^{\rm (P)}$, and by the standard ket symbol in
the other two Hilbert spaces, $|\psi\kt \in {\cal H}^{\rm (F,S)}$. The
two spaces ${\cal H}^{\rm (F,S)}$ are chosen as identical when only
their ket elements are being considered. These two spaces are
identical as vector spaces which should  be denoted by a~dedicated
symbol ${\cal V}^{\rm (F)}={\cal V}^{\rm (S)}$. They only dif\/fer in the
respective def\/inition of their respective linear functionals, i.e.,
of the bra-vector elements of the dual vector spaces marked by the
prime, ${{\cal V}^{\rm (F)}}' \neq {{\cal V}^{\rm (S)}}'$. The
correspondence  between the three dual vector spaces can be
represented by the following diagram,
 \ben
  \ba
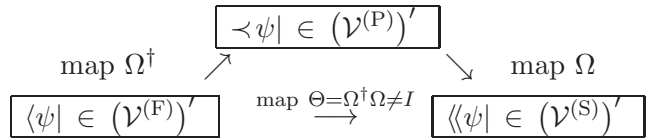
    \begin{array}{|c|}
 \hline%
  \pbr \psi|\,\in\,
 \left ({\cal V}^{\rm (P)}\right )'\
 \\
 \hline
 \ea
\\  {\rm map}\
 \Omega^\dagger \ \ \ \
 \nearrow \ \  \  \ \ \ \ \ \
 \ \ \ \ \ \ \ \
\ \ \ \ \  \searrow \ \ \  {\rm map}\ \Omega \\
 \begin{array}{|c|}
 \hline%
  \br \psi|\,\in\,
 \left ({\cal V}^{\rm (F)}\right )'\ \\
  \hline
 \ea\ \ \
 \stackrel{ {\rm map}\ \Theta=\Omega^\dagger\Omega\neq I }{ \longrightarrow }
 \
 \begin{array}{|c|}
 \hline%
 \bbr \psi|\,\in\,
 \left ({\cal V}^{\rm (S)}\right )'\
 \\
 \hline
 \ea
\\
\ea
 \een
In this notation it is easy to def\/ine $\bbr \Psi|=\br \Psi|\Theta$
or to work with the map $\Omega: |\psi_n\kt \to |\psi_n\pkt$. It is
equally easy to verify the unitarity of the map between the Hilbert
spaces ${\cal H}^{\rm (P)}$ and ${\cal H}^{\rm (S)}$,
 \ben
 \pbr \psi|\phi\pkt = \bbr \psi|\phi\kt .
 \een
As long as $({\cal V}^{\rm (F)})'\neq ({\cal V}^{\rm (S)})'$ we must
distinguish between the explicit def\/initions of the respective
operations of Hermitian conjugation. The operation marked by
single-cross superscript $^\dagger$ in ${\cal H}^{\rm (F)}$ {\em must be
distinguished} from the one active in ${\cal H}^{\rm (S)}$ and marked by
double-cross $^\ddagger$.

The acceptance of the notation recommended in this section enables
us to introduce a~modi\-f\/ied bra symbol in ${\cal H}^{\rm (S)}$,
$(|\Psi\kt )^\ddagger = \bbr \Psi|  \in   ({\cal V}^{\rm (S)})'$ while
keeping the traditional $(|\Psi\kt )^\dagger = \br \Psi|   \in
({\cal V}^{\rm (F)})'$ unchanged. There is no doubt that the use of
three Hilbert spaces ${\cal H}^{\rm (P,F,S)}$ does not contradict any
principle of textbook quantum mechanics. Indeed, all the {\em
physical} operators of observables remain Hermitian in ${\cal
H}^{\rm (P)}$ while the other two spaces ${\cal H}^{\rm (F,S)}$ are just
auxiliary.

\section{Scattering and problems with long-ranged non-Hermiticities}\label{problems}

The variability of $\Theta$s and/or $\Omega$s could inspire
optimistic expectations concerning the extension of the
cryptohermitian description of physics to scattering scenario.
Unfortunately, several se\-rious obstacles were found and formulated
by Jones~\cite{Jones}. He argued that Dyson maps $\Omega$ must be
necessarily long-ranged leading to an apparent violation of
causality in scattering (cf.\ Section~\ref{oJones} below). A return
to optimism has been initiated in~\cite{prd} where some of the
most striking dif\/f\/iculties were circumvented via a replacement of
the current, strictly local interactions~$V(x)$ by their minimally
nonlocal alternatives (cf.\ Section~\ref{oprd} below). The f\/irst
fully satisfactory model of scattering has been found in~\cite{FT} (cf.\ Section~\ref{oFT} below). A continuation of these
developments will be reported here immediately after a brief summary
of older results.

\subsection{The f\/irst problem: The causality-violating waves in the
scattering \label{oJones} }

In numerous recent applications of the above-mentioned three-space
representation of quantum theory to bound states a conf\/lict was
encountered between the friendliness of the calculations in
unphysical ${\cal H}^{\rm (F)}$ and the unpleasant complications arising
during the transition to the two alternative physical spaces ${\cal
H}^{\rm (S,P)}$. Many concrete Hamiltonians $H$  happened to possess a
particularly simple form in space ${\cal H}^{\rm (F)}$ and vice versa.

The transition to the correct space may be perceived as a fairly
unnatural operation. The more so if one tries to describe the
non-Hermitian scattering \cite{Cannata,Jonesdva}. For these reasons,
the counterintuitive character of the textbook Hermiticity of $H$ in
unfriendly ${\cal H}^{\rm (S)}$ is ref\/lected by the terminology in which
the cryptohermitian operators are nicknamed quasi-Hermitian rather
than Hermitian \cite{Geyer,Dieudonne}.

The main dif\/ference between the two Hilbert spaces ${\cal
H}^{\rm (F,S)}$ concerns their inner products between elements
$|\psi\rangle$ and $|\phi\rangle$. In the most usual coordinate
representation language we have
 \be
 \langle \psi|\phi \rangle^{\rm (F)}=
 \int \psi^*(x) \phi(x)\, dx=\langle \psi|\phi \rangle
 \qquad {\rm in}
 \quad {\cal H}^{\rm (F)},
 \label{diracs}
 \ee
and
 \be
 \langle \psi|\phi \rangle^{\rm (S)}=
 \iint \psi^*(x) \Theta(x,x') \phi(x')\, dx dx'=
 \langle \psi|\Theta|\phi \rangle
 \qquad {\rm in}
 \quad {\cal H}^{\rm (S)},
 \label{e777}
 \ee
where $\Theta(x,x') \neq \delta(x-x')$. The consequences of the
transition from equation~(\ref{diracs}) to equation~(\ref{e777}) are
nontrivial. For example, according to Mostafazadeh \cite{delta} the
assumption of having a short-range and strictly local $V=V(x)$ leads
immediately to the necessity of using a strongly nondiagonal metric
kernel $\Theta(x,x')$ in equation~(\ref{e777}).

The latter long-range nonlocality is easily visualized in
Runge--Kutta picture where the coordinates are represented by a
lattice of discrete points $x=x_j$, $j = \ldots, -1, 0, 1, \ldots$
and where all the dif\/ferences $x_{j+1}-x_j=h$ are the same. In this
approximation the typical doubly inf\/inite matrix
$\Theta_{i,j}=\Theta(x_i,x_j)$  is dominated by the unperturbed unit
matrix $\Theta^{(0)}_{i,j}=\delta_{i,j}$. The non-negligible
correction given, e.g., by equation~(16) in \cite{Jones} will be strongly
non-diagonal. The metric can comfortably be expanded in the sum of
elementary matrices, $\sum_k e^{-\beta h  k}\Theta^{(k)}_{i,j}$
where each coef\/f\/icient is sparse and strongly non-diagonal,
 \be
 \Theta^{(1)}=
 \begin{array}{|cccc|cccc|}
 \hline
 \ddots &&&&&&& _. . ^.\\
 &1&&&&&1&\\
 &&1&&&1&&\\
 &&&1&1&&&\\
 \hline
 &&&1&1&&&\\
 &&1&&&1&&\\
 &1&&&&&1&\\
 _. . ^. &&&&&&& \ddots \\
 \hline
 \ea \,,
 \ \
 \Theta^{(2)}=
 \begin{array}{|cccc|cccc|}
 \hline
 &\ddots &&&&& _. . ^.&\\
 \ddots &&1&&&1&&_. . ^.\\
 &1&&1&1&&1&\\
 &&1&&&1&&\\
 \hline
 &&1&&&1&&\\
 &1&&1&1&&1&\\
 _. . ^.&&1&&&1&&\ddots \\
 & _. . ^. &&&&& \ddots &\\
 \hline
 \ea\,,\ \ldots.
 \label{orange}
 \ee
These cross-shaped matrices couple remote coordinates $x_i$ and
$x_j$. A conf\/lict emerges between our use of the common Hilbert
space ${\cal H}^{\rm (F)}$ (which keeps trace of the intuitive principle
of correspondence) and of its amendment ${\cal H}^{\rm (S)}$ (only there
the correct probabilistic interpretation of the observables is
achieved). The core of the problem lies in the manifest loss of the
concept of the asymptotically free motion. In words of~\cite{Jonesdva} one has to conclude that the
causality-violating nature of metric is certainly ``changing the
physical picture drastically''.

\subsection{The second problem:  Spatial symmetry violations
\label{oprd} }

\looseness=1
The f\/irst steps towards the desirable return to the standard picture
of scattering induced by the Hamiltonian $H$ which proves
non-Hermitian in space ${\cal H}^{\rm (F)}$ were proposed in our
comment~\cite{prd}. We tried to resolve there those of the paradoxes
covered by paper~\cite{Jonesdva} which were available prior to
publication \cite{London}. In particular we paid attention to the
apparently unavoidable presence of causality-violating waves which
seemed to emerge in the right spatial inf\/inity. In the context of
the specif\/ic non-Hermitian delta-function scattering models this
violation of causality seems to result from the strict locality of
$V=V(x)$. We concluded that such an assumption proves too strong and
that it must be weakened for the given purpose.

In a detailed discussion of the causality-violation paradox  we
employed the same discretization. We assumed that in Hamiltonians
$H=-d^2/dx^2+{V}$ the potential of any form must be combined with
the  tridiagonal Runge--Kutta version of the kinetic-energy operator,
 \be
 H= -\triangle
 + V,\qquad
 -\triangle=
 \begin{array}{|cccccc|}
 \hline
 \ddots &\ddots &&&& \\
 \ddots &2&-1&&&\\
 &-1&2&-1&&\\
 &&-1&2&-1&\\
 &&&-1&2&\ddots \\
  &&&&\ddots & \ddots \\
 \hline
 \ea\,.
 \label{dvanact}
 \ee
This helped us to simplify some technicalities and to clarify the
way in which the asymptotically non-vanishing nondiagonality in
$\Theta_{i,j}$ is generated by the diagonality of potentials
$V=V(x_i)$. We then decided to consider only such families of
Hamiltonians for which the unpleasant non-diagonality of the metric
disappears,
 \be
 \Theta(x_j,y_k)
  \approx  c(x_j) \delta_{j,k},
 \qquad |x_j| \gg 1,\qquad |y_k|\gg 1
 \label{assu}
 \ee
{\sloppy (cf.\ equation~(23) in \cite{prd}). Then we only had to recollect that the
(crypto) Hermiticity of~$H$ in~${\cal H}^{\rm (S)}$ is a condition which
acquires the utterly elementary form (\ref{crypto}) in ${\cal
H}^{\rm (F)}$. As a~linear-equation constraint imposed upon the matrix
elements of metric $\Theta$ this equation can serve as~an~independent, nonperturbative source of information about the
nonlocalities in met\-rics~(\ref{orange}).\looseness=1

}

In~\cite{prd} we decided to shorten the range of the inf\/luence
of the non-Hermiticity. After the insertion of ansatz (\ref{assu})
in equation~(\ref{crypto}) a series of our algebraic trial and error
experiments revealed that in order to achieve a certain internal
consistency of our requirements we might replace the usual complex
and diagonal non-Hermitian matrix $V(x_i)$ in $H=-\triangle+ V(x)$
by its two-diagonal real and non-Hermitian analogue of the form
 \be
 V^{(a,b,c,\ldots )}=\left [\begin {array}{cccc|cccc}
 \hline
 &\ddots&{}&{}&{}&{}&{}&{}{}{}{}\\
 \ddots&{}& -c &{}&{}&{}&{}&{}{}{}{}\\
 &c&{}&-b&{}&{}&{}&{}{}{} \\
 &{}&b&{}&-a&{}&{}&{}{}{}{}\\
 \hline
 {}&{}&{}&a&{}&-b&{}&{}{} \\
 {}&{}&{}&{}&b&{}&-c&{}{}{}\\
 {}  &{}&{}&{}&{}&c&{}&\ddots{}{}\\
 {}&{}&{}&{}&{}&{}&\ddots&{}\\
 \hline
 \end {array}\right ] .
 \label{muli}
 \ee
A multiparametric Schr\"{o}dinger Hamiltonian $H=-\triangle+
V^{(a,b,c,\ldots )}$ has been found which is Hermitian in the
Hilbert space ${\cal H}^{\rm (S)}$ where the exact metric operator has
the following nontrivial but still fully diagonal form,
 \be
 \Theta^{(a,b,c,\ldots )}=\left [\begin {array}{cccc|cccc}
 \hline
 \ddots&&{}&{}&{}&{}&{}&{}{}{}{}\\
 &{\theta_{-5}}&  &{}&{}&{}&{}&{}{}{}{}\\
 &&{\theta_{-3}}&&{}&{}&{}&{}{}{} \\
 &{}&&{\theta_{-1}}&&{}&{}&{}{}{}{}\\
 \hline
 {}&{}&{}&&{\theta_1}&&{}&{}{} \\
 {}&{}&{}&{}&&{\theta_3}&&{}{}{}\\
 {}  &{}&{}&{}&{}&&{\theta_5}&{}{}\\
 {}&{}&{}&{}&{}&{}&&\ddots\\
 \hline
 \end {array}\right ] .
 \label{formul}
 \ee
Matrix elements are explicitly known,
 \begin{gather*}
 \theta_{\pm 1}=(1\pm a)(1-b^2)(1-c^2)(1-d^2) \cdots,
\\
 \theta_{\pm 3}=(1\pm a)(1\pm b)^2(1-c^2)(1-d^2) \cdots ,
\\
 \theta_{\pm 5}=(1\pm a)(1\pm b)^2(1\pm c)^2(1-d^2) \cdots.
 \end{gather*}
We separated the ``in'' and ``out'' solutions not only in ${\cal
H}^{\rm (F)}$ but also in ${\cal H}^{\rm (S)}$. A causality-observing
physical picture of scattering was demonstrated to exist.
Unfortunately, the amendment of the potentials proved incomplete. We
did not manage to remove another shortcoming of equation~(\ref{formul})
where, for a generic set of parameters $a, b, \ldots$, the
asymptotic measure $c(x_j)-1$ of the anomaly of the f\/lux in
equation~(\ref{assu}) remains long-ranged (cf.\ a footnote in~\cite{Jonesdva}).

\subsection[Problems resolved: ${\cal PT}$-symmetric models of scattering of~\cite{FT}]{Problems resolved: $\boldsymbol{{\cal PT}}$-symmetric models of scattering of~\cite{FT}}\label{oFT}

Having accepted the above-cited words of critique we sought for a
removal of the long-range anomalies. In letter~\cite{FT} we made
another step towards a fundamental theory of scattering based on the
Hamiltonians which appear non-Hermitian in space ${\cal H}^{\rm (F)}$.
Another, amended family of Runge--Kutta Hamiltonians has been found
there preserving the kinetic + potential-energy structure, $H = T +
V$. We restricted our attention to the following two-diagonal matrix
interaction proportional to real coupling parameter $g$ and
mimicking the existence of two interaction centers at a distance
$\sim 2M$,
 \be
 V^{(g)}=\left [\begin {array}{cc|cc|cc|cc|cc}
 \hline
 &\ddots&{}&{}&{}&{}&{}&{}&{}&{}{}\\
 \ddots&{}& 0 &{}&{}&{}&{}&{}&{}&{}{}\\
 \hline
 &0&{}&-g&{}&{}&{}&{}&{}&{} \\
 &{}&g&{}&0&{}&{}&{}&{}&{}{}\\
 \hline
 &{}&{} &0&{}&\ddots&{}&{}&{}&{}{}\\
 &{}&{} &&{\ddots}&&{0}&{}&{}&{}{}\\
 \hline
 &{}&{}&{}&{}&0&{}&g&{}&{}{} \\
 &{}&{}&{}&
 \multicolumn{2}{c|}{\underbrace{\ \ \ \ \ \ \ \ \ \ \ \ }}
 &-g&{}&0&{}{}{}\\
 \hline
  & &   &&
 \multicolumn{2}{c|}{   2M-3 }
 & &0& &\ddots  \\
  & &   &&
 \multicolumn{2}{c|}{    {\rm \ columns }}
  & & & \ddots&  \\
 \hline
 \end {array}\right ] .
 \label{mode}
 \ee
The key point was that we achieved an asymptotic spatial symmetry of
the metric operator,
 \be
 c(x_j)= c(-x_j)=1 ,\qquad |x_j| \gg 1 .
 \label{asympto}
 \ee
The overall picture of\/fered by equation~(\ref{asympto}) looks
satisfactory, in the discrete-coordinate approximation at least. The
metric assigned to our Hamiltonian $H$ remains diagonal, even though
it is still dif\/ferent from the unit matrix. We can reasonably expect
that for the majority of interactions the diagonality of this matrix
may easily happen to lead just to a smooth change of the measure in
equation~(\ref{diracs}) giving just the replacement $ dx \to d\mu(x)$ in
the limit $h\to 0$. In spite of having a nontrivial Hilbert space,
the changes due to $\Theta \neq I$ may remain inessential and the
limiting metric $\Theta$ may be called ``local''.

\looseness=1
In contrast, whenever we relax the constraint of diagonality and
admit a $(2R-1)$-diagonal metric $\Theta$ with $R\geq 2$, the
decrease of the discretization length $h\to 0$ may lead, in gene\-ral,
to a matrix $\Theta$ which can be, say, a polynomial function of the
kinetic energy operator~$\triangle$ of equation~(\ref{dvanact}). The
$h\to 0$ limit of the discrete inner product would not lead to the
mere double-integration formula (\ref{e777}) but rather to a very
complicated (e.g., integro-dif\/ferential) recipe.

In our present text we intend to omit the discussion of all the
similar mathematical subtleties, emphasizing only that they might
play an important role even in the present class of illustrative
examples. This is also the reason why we call our band-matrix
metrics quasilocal. In the limit $h\to 0$, their range will
certainly shrink to zero but the same feature characterizes also the
kinetic-energy operator $\triangle$.

\section{Band-matrix metrics:  their meaning and construction}\label{core}

The fact that we can only reconstruct the reality from our
discretized scattering models via the limiting transition $h\to 0$
has an important constructive aspect. Let us recollect that the core
of the consistence of the theory at $h=0$ lies in a reconciliation
of the non-Hermiticity of dif\/ferential equation~(\ref{SEor}) in ${\cal
H}^{\rm (F)}$ with the asymptotic locality of boundary conditions
(\ref{scatbc}). We saw that at $h>0$ both these problems found their
resolution in~\cite{FT} where $H$ was simply re-interpreted as
Hermitian in ${\cal H}^{\rm (S)}$ and where the diagonality of the
metric at $h>0$ of\/fered an easy guarantee of its locality in the
limit $h=0$.

Now we have to add that the classif\/ication of possibilities as
presented in~\cite{FT} is incomplete because we required there
the strict locality of the metric. On this background we may brief\/ly
characterize our present new results as a broadening of the picture
of unitary scattering as of\/fered in~\cite{FT}.

\looseness=1
Our present key idea is that the standard form of the asymptotic
boundary conditions for scattering (def\/ined, naturally, in terms of
plane waves) would not be lost even if we assume that our metric
$\Theta$ becomes weakly non-diagonal. For example, certain constant
tridiagonal matrices $\Theta$ constructed at a given lattice
constant $h>0$ may appear as proportional to the Laplace operator
(i.e., to the kinetic-energy operator) in the continuous coordinate
limit \mbox{$h \to 0$}.

This point of view can be generalized to all of the
$(2R-1)$-diagonal matrices of the metric~$\Theta$ where the integer
$R$ remains f\/ixed during the decrease of $h\to 0$. The resulting
limits $\Theta$ may often prove expressible in terms of the powers
of the Laplace operator so that we may call all of these $h=0$
metrics quasilocal.

As a consequence, a broad family of new metrics $\Theta$ could be
obtained in the limit $h=0$. They would still, by construction,
leave the scattering unitary. The resulting picture of physics will
vary with the choice of their structure at $h>0$. In order to be
able to understand this situation we would need to be able to
construct the $(2R-1)$-diagonal matrices $\Theta$ in closed form at
any $h>0$.

This possibility has been ignored before. Let us now f\/ill this gap
via our solvable non-Hermitian example.

\subsection{The choice of the model}

Let us select the class of interaction models (\ref{mode}) of~\cite{FT} and, for illustration purposes, let us pick up the
special case where, formally, $2M-3=-1$,
 \be
 H= \left[ \begin {array}{ccccccc} &\ddots&{}&{}&{}&{}&{}\\{}\ddots&2
&-1&{}&{}&{}&{}\\{}{}&-1&2&-1-{g}&{}&{}&{}\\{}{}&{}
&-1+{g}&2&-1+{g}&{}&{}\\{}&{}&{}&-1-{g}&2&-1&{}
\\{}{}&{}&{}&{}&-1&2&\ddots\\{}{}&{}&{}&{}&{}&\ddots&\ddots
\end {array} \right] .
\label{slef}
 \ee
{\sloppy \looseness=1 This is a tridiagonal, doubly inf\/inite Hamiltonian matrix  which can
be complemented by a discrete version of asymptotic boundary
conditions (\ref{scatbc}) for scattering wave functions. The details
including the explicit formulae for the ref\/lection coef\/f\/icient
$R(E)$ and for the transmission coef\/f\/icient $T(E)$ may be found in~\cite{FT} showing that $|T(E)|^2+|R(E)|^2=1$.  This means that
the standard probability f\/low is conserved (i.e., the process is
unitary) even when the scattering is controlled by our non-Hermitian
Hamilto\-nian~(\ref{slef}).

}

The unitarity of the model is related to the existence of a metric
which is asymptotically diagonal. At any nonvanishing $h>0$ we may
easily f\/ind the fully diagonal metric containing just a single
anomalous matrix element,
 \be
 \Theta=\Theta_1=
 \left[ \begin {array}{ccccc} {}\ddots&{}&
&{}&{}\\
{}{}&1&{}&&{}\\
{}{}&{}& {(1+{g})}/{(1-{g})}&{}&
\\
{}{}&{}&{}&1&{}\\
{}{}&{}&{}&{}&\ddots
\end {array} \right] .
\label{metr1}
 \ee
Having worked in discrete coordinate representation we may conclude
that our matrix $\Theta_1$ commutes with the usual diagonal operator
of the coordinate. The anomalous matrix element
$z_1=(1+{g})/(1-{g})=1+2{g}/(1-{g})$ degenerates to 1 in the
no-interaction limit (i.e., Hermitian limit) $\ g\to 0$. One should
notice that the spectral singularity \cite{spectral} emerges  at
${g}=\pm 1$.

\subsection[The existence of band-matrix metrics with $2R-1\leq 9$ diagonals]{The existence of band-matrix metrics with $\boldsymbol{2R-1\leq 9}$ diagonals}

{\sloppy \looseness=1 Our choice of the elementary toy-model Hamiltonian (\ref{slef})
facilitates the constructive search for the alternatives to the
diagonal metric (\ref{metr1}). Indeed, whenever we perceive the
class of admissible metrics $\Theta=\Theta(H)$ as composed of {\em
all} the possible Hermitian, positive and invertible matrices which
are compatible with the Hamiltonian-dependent quasi-Hermiticity
constraint (\ref{crypto}) we may (and shall) take this equation
simply as a linear algebraic system of (inf\/initely many) equations
which are to be satisf\/ied by our pair of matrices $H$ (which is
given in advance) and $\Theta$ (which is to be specif\/ied as one of
solutions of equation~(\ref{crypto})).

}

{\sloppy \looseness=1
There exist several technical obstacles which make such a project
not entirely trivial. Firstly, we have to employ some ``manual"
linear algebra in order to reduce the inf\/inite-dimensional
equation~(\ref{crypto}) to its f\/inite-dimensional subsystem. By the trial
and error techniques we succeeded in revealing that such a reduction
may even preserve many parallels with  diagonal $\Theta_1$. In this
spirit we decided to search for certain  doubly-inf\/inite-matrix
solutions $\Theta_R$ which exhibit a very specif\/ic band-matrix
property $\left (\Theta_R\right )_{k,m}=0$ whenever $|k-m|>R$.
Empirically we  revealed that each such sparse, $(2R-1)$-diagonal
mat\-rix solution $\Theta_R$ may, furthermore, be required to contain
solely $R$ nonvanishing diagonals. This possibility is supported by
the tridiagonal matrix ansatz for $\Theta_R$ at $R=2$ gi\-ving
 \ben
 \Theta_2=
 \left[ \begin {array}{ccccccc} {}&\ddots&{}&{}&{}&{}&{}\\{}\ddots&{\ddots}&1
&{}&{}&{}&{}\\{}{}&1&{0}&1+{g}&{}&{}&{}\\{}{}&{}&1+{g}&{0}&1+{g}
&{}&{}\\{}{}&{}&{}&1+{g}&{0}&1&{}\\{}{}&{}&{}&{}&1&{\ddots}&\ddots
\\{}{}&{}&{}&{}&{}&\ddots&{}\end {array} \right] .
 \een
A reconf\/irmation is found at $R=3$,} with
 \[
 \Theta_3=
\left[ \begin {array}{ccccccccc}
{\ddots}&{\ddots}&{\ddots}&&{}&{}&{}&{}&{}{}\\
{\ddots}&1&{0}&1&{}&{}&{}&{}&{}{}\\
\ddots&0&1&{0}&1+{g}&{}&{}&{}&{}{}\\
&{}{}1&{0}&1-{{g}}^{2}&{0}&1-{{g}}^{2}&{}&{}&{}{}
\\
&{}{}{}&1+{g}&{0}&\dfrac{\left( 1+{g} \right)  \left(
1-2\,{{g}}^{2}
\right)}{1-{g}}&{0}&1+{g}&{}&{}\\
{}&{}&{}&1-{{g}}^{2} &{0}&1-{{g}}^{2}&{0}&1{}&\\
&{}{}{}&{}&{}&1+{g}&{0}&1&0&\ddots{}
\\
&{}{}{}&{}&{}&{}&1&0&1&{{\ddots}}\\
{}{}{}&{}&{}&{}&&{}&{\ddots}&{{\ddots}}&{{\ddots}}
\end {array} \right]
 ,
 \]
and at $R=4$, with
 \ben
 \Theta_4=
 \left[ \begin {array}{ccccccccccc}
 &{{{{}} \bf \ddots}}&&{{{{}} \bf \ddots}}&&&&&&&\\
 {{{{}} \bf \ddots}}& {{{}} \bf \ddots}&1&{{{}} \bf \ddots}&{\rm {{}} 1}&&&&&&\\
 &1 &0&{\rm {{}} 1}&0&{a}&&&&&\\
 {{{}} \bf \ddots}&{{{}} \bf \ddots}&{\rm {{}} 1}&{{{}} \bf \ddots}
    &b &{{{}} \bf \ddots}&{b}&&&&\\
 &{\rm {{}} 1}&0&b&&c&&{b}&&&\\
 &&{a}&{{{}} \bf \ddots}&c&&c &{{{}} \bf \ddots}&{a}&&\\&&
 &{b}&&c&&b&0&{\rm {{}} 1}&\\&&
 &&{b}&{{{}} \bf \ddots}&b&{{{}} \bf \ddots}&{\rm {{}} 1} &{{{}} \bf \ddots}&{{{}} \bf \ddots}\\&
 &&&&{a}&0&{\rm {{}} 1}&0&1&\\&
 & &&&&{\rm {{}} 1}&{{{}} \bf \ddots}&1&{{{}} \bf \ddots}&
 {{{{}} \bf \ddots}}\\&
 &&&&&&{{{}} \bf \ddots}&&{{{{}} \bf \ddots}}&
 \end {array} \right],
 \een
where we abbreviated
 $
 a = 1+{g}$,
 $b=1-{g}^2$ and
 $c=(1+{g})(1-2{g}^2)$. Finally, we also solved equation~(\ref{crypto})
with the similar $R=5$ ansatz
 \ben
 \Theta_5=\left[ \begin {array}{ccccccccccccc} {\ddots}&{}&{\ddots}&{}&\ddots
 &{}&{}&{}&{}&{}&{}&{}&{}
\\{}{}&{1}&{}&1&{}&1&{}&{}&{}&{}&{}&{}&{}\\{}{\ddots}&{}&1
&{}&1&{}&a&{}&{}&{}&{}&{}&{}\\{}{}&1&{}&1&{}&b&{}&b&{}&{}&{}&{}&{}
\\{}\ddots&{}&1&{}&b&{}&c&{}&b&{}&{}&{}&{}\\{}{}&1&{}
&b&{}&{d}&{}&{d}&{}&b&{}&{}&{}\\{}{}&{}&a&{}&c&{}&z_5&{}&c&{}&a&{}&{}
\\{}{}&{}&{}&b&{}&{d}&{}&{d}&{}&b&{}&1&{}\\{}{}&{}&{}
&{}&b&{}&c&{}&b&{}&1&{}&\ddots\\{}{}&{}&{}&{}&{}&b&{}&b&{}&1&{}&1&{}
\\{}{}&{}&{}&{}&{}&{}&a&{}&1&{}&1&{}&{\ddots}\\{}{}&{}&{}
&{}&{}&{}&{}&1&{}&1&{}&{1}&{}\\{}{}&{}&{}&{}&{}&{}&{}&{}
&\ddots&{}&{\ddots}&{}&{\ddots}
\end {array} \right]
 \een
and we obtained, again, $a=1+{g}$, $b=1-{g}^2$,
$c=(1+{g}) (1-2\,{g}^2)$, ${d}=(1-{g}^2) (1-2 {g}^2)$ and
 \ben
 z_5= {\frac { \left( 1+{g} \right)
  \left( 1-2\,{{g}}^{2} \right) ^{2}}{1-{g}}} .
 \een

\subsection[The extrapolation and construction of all the metrics $\Theta_R$]{The extrapolation and construction of all the metrics $\boldsymbol{\Theta_R}$}

As a result of our computer-assisted exact and systematic
reconstruction of the metrics $\Theta=\Theta_k$ compatible with
Hamiltonian $H$ via quasi-Hermiticity condition (\ref{crypto}) we
arrive at the following extrapolation of our $k\leq 5$ rigorous
symbolic-manipulation results to any integer $k>5$.
\begin{itemize}\itemsep=0pt

\item
 By construction, each metric $\Theta_k$
possesses the form of a doubly inf\/inite and symmetric
$(2k-1)$-diagonal real matrix with $k$ nonvanishing diagonals
interlaced by $k-1$ zero diagonals.

\item
 Up to their central $k$-plets, all the matrix elements of
$\Theta_k$ along each non-vanishing diagonal are equal to one.

\item
 If we assume $k\geq 2$ and abbreviate $a = 1+{g}$ and $b=1-{g}^2$
the leftmost non-trivial $k$-plet  $(a,b,b,\ldots,b,b,a)$ of
non-unit matrix elements of our metric $\Theta_k$ contains two
boundary $a$'s complemented by the $(k-2)$-plet of $b$'s.

\item
 After we abbreviate  $c=(1+{g})\,(1-2\,{g}^2)$ and
${d}=(1-{g}^2) (1-2 {g}^2)$ and assume that $k\geq 4$, we obtain
the subsequent $k$-plet in the explicit form
$(b,c,d,d,\ldots,d,d,c,b)$ f\/illed by $k-4$ $d$'s.

\item
 With  $e=(1+{g}) (1-2 {g}^2)^2$ and
${f}=(1-{g}^2) (1-2 {g}^2)^2$ and  $k\geq 6$ we get the next
$k$-plet $(b,d,e,f,f,\ldots,f,f,e,d,b)$ containing $k-6$ $f$'s.
Etc.

\end{itemize}

The general pattern is obvious: the quadruplets of matrix elements
sitting in the four (viz., ``North'', ``East'', ``South'' and ``West'')
corners may be ordered in an inwards-running sequence of vertices $
a=A_1$, $c=A_2$, $e=A_3$, $\ldots$ (cf.\ Table~\ref{pexp4}). These values
are complemented by the related multiplets of the wedge-f\/illing
elements $ b=B_1$, $d=B_2$, $f=B_3$, $\ldots$. At the odd subscripts there
emerges an anomalous, non-polynomial central matrix element $z_k$
equal to fractions $a/(1-{g})$, $c/(1-{g})$, $e/(1-{g})$, $\ldots$
at $k=1,3,5,\ldots$, respectively.

The existence of the above extrapolation rules enables us to
formulate a simplif\/ied ansatz for the next matrix $\Theta_R$ at
$R=6$,
 \ben
 \Theta_6=\left[ \begin {array}{ccccccccccccccc}
 & {\ddots}&{}&{\ddots}&{}&{\ddots}&&{}&{}&{}&{}&{}&{}&{}
 &\\{\ddots}&
  {}&{1}&{}&{1}&{}&1&{}&{}&{}&{}&{}&{}&{}
 &\\&
1&{}&{1}&{}&1&{}&a&{}&{}&{}&{}&{}&{}
 &\\{\ddots}&
{}{}&{1}&{} &1&{}&b&{}&b&{}&{}&{}&{}&{}
 &\\&
1&{}&1&{}&b&{}&c&{}&b&{}&{}&{}&{}
 &\\{\ddots}&
{}{}&1&{}&b&{}&d&{}&d&{}&b&{}&{}&{}
 &\\&
1&{}&b &{}&d&{}&e&{}&d&{}&b&{}&{}
 &\\&
{}{}&a&{}&c&{}&e&{}&e&{}&c&{}&a&{}
 &\\&
{}{}&{}&b&{}&d&{}&e&{}&d&{}&b&{}&1
 &\\&
{}{}&{}&{} &b&{}&d&{}&d&{}&b&{}&1&{}
 &{\ddots}\\&
{}{}&{}&{}&{}&b&{}&c&{}&b&{}&1&{}&1
 &\\&
{}{}&{}&{}&{}&{}&b&{}&b&{}&1&{}&{1}&{}
 &{\ddots}\\&
{}{}&{}&{} &{}&{}&{}&a&{}&1&{}&{1}&{}&1
 &\\&
{}{}&{}&{}&{}&{}&{}&{}&1&{}&{1}&{}&{1}&{}
 &{\ddots}\\&
{}{}&{}&{}&{}&{}&{}&{}&&{\ddots}&{}&{\ddots}&{}&{\ddots}&
\end {array} \right] .
 \een
It passes the test after insertion in equation~(\ref{crypto}). The
tedious computer-assisted direct solution becomes replaced by an
elementary verif\/ication of the choice of $e= \left( 1+{g} \right)
\left( 1-2 {{g}}^{2} \right)^{2}$.

\begin{table}[t]
\caption{Matrix elements of the metrics $\Theta_R$.} \label{pexp4}

\vspace{2mm}

\centering
\begin{tabular}{||c|l|l|l||}
\hline \hline
\tsep{1ex}\bsep{1ex}  $R$ & {\rm central element}&{\rm corner element}&{\rm wedge element}\\
 \hline \hline
\tsep{1ex}\bsep{1ex} 1&$z_1={A}_1/(1-g)$& --- & ---
 \\
 \bsep{1ex}3&$z_3={A}_2/(1-g)$& $a={A}_1=(1+g)$& $b={B}_1=(1-g^2$)
 \\
\bsep{1ex} 5&$z_5={A}_3/(1-g)$ & $c={A}_2=(1+g)(1-2g^2)$ & $d={B}_2=(1-g^2)(1-2g^2)$
 \\
\bsep{1ex} 7&$z_7={A}_4/(1-g)$& $e={A}_3=(1+g)(1-2g^2)^2$ & $f={B}_3=(1-g^2)(1-2g^2)^2$
 \\
\bsep{1ex} 9&$z_9={A}_5/(1-g)$& ${A}_4=(1+g)(1-2g^2)^3$& ${B}_4=(1-g^2)(1-2g^2)^3$
 \\
\bsep{1ex} {$\vdots$}
 &\multicolumn{1}{c|}{$\vdots$}
 &\multicolumn{1}{c|}{$\vdots$}
 &\multicolumn{1}{c||}{$\vdots$}
 \\
\bsep{1ex} $2k+1$& $z_{2k+1}={A}_{k+1}/(1-g)$& ${A}_k=(1+g)(1-2g^2)^{k-1}$
 &${B}_k=(1-g^2)(1-2g^2)^{k-1}$
 \\
 \hline \hline
\end{tabular}
\end{table}

 \looseness=2
In a climax of our analysis let us now complement our algorithmic
set of extrapolation rules by their rigorous proof. For this purpose
we may select an odd integer $R=2k+1$ and reinterpret
Table~\ref{pexp4} as a recurrent pattern (i.e., the proof proceeds
by mathematical induction). The corner element $A_k$ is most easily
deduced from the previous line since $A_k=(1-g)z_{2k-1}$ and only
the values of $B_k$ and $z_{2k+1}$ must be deduced from the
quasi-Hermiticity condi\-tion~(\ref{crypto}) rewritten as a doubly
inf\/inite matrix set of linear equations using the input Hamilto\-nian~(\ref{slef}),
 \begin{gather}
 \big [H^\dagger \Theta_{2k+1}- \Theta_{2k+1} H
 \big ]_{mn}=0 ,\qquad
 m,n = 0, \pm 1, \ldots .
  \label{quas}
 \end{gather}
This task is facilitated by the tridiagonality of $H$ and it may be
also guided by our last illustrative sparse-matrix ansatz for
$\Theta_R$ at the next odd $R=7$,
 \ben
 \Theta_7=
 \left[\begin {array}{ccccccccccccccccc}
 {}&{}&{}&{}&{{\ddots}}&{}&{\ddots}&{}&{}&{}&{}&{}&{}&{}&{}&{}&{}
 \\[-1.8mm] {{}}{}&{}&{}&{{\ddots}}&{}&1&{}&1&{}&{}&{}&{}&{}&{}&{}&{}&{}
 \\[-1.8mm]{{}}{}&{}&{{\ddots}}&{}&1&{}&1&{}&a&{}&{}&{}&{}&{}&{}&{}&{}
 \\[-1.8mm]{{}}{}&{{\ddots}}&{}&1&{}&1&{}&b&{}&b&{}&{}&{}&{}&{}&{}&{}
 \\[-1.8mm]{\ddots}&{}&1&{}&1&{}&b&{}&c&{}&b&{}&{}&{}&{}&{}&{}
 \\[-1.8mm]{{}}{}&1&{}&1&{}&b&{}&d&{}&d&{}&b&{}&{}&{}&{}&{}
 \\[-1.8mm]{\ddots}&{}&1&{}&b&{}&d&{}&{\it e}&{}&d&{}&b&{}&{}&{}&{}
 \\[-1.8mm]{{}}{}&1&{}&b&{}&d&{}&f&{}&f&{}&d&{}&b&{}&{}&{}
 \\[-1.8mm]{{}}{}&{}&a&{}&c&{}&{\it e}&{}&z_7&{}&{\it e}&{}&c&{}&a&{}&{}
 \\[-1.8mm]{{}}{}&{}&{}&b&{}&d&{}&f&{}&f&{}&d&{}&b&{}&1&{}
 \\[-1.8mm]{{}}{}&{}&{}&{}&b&{}&d&{}&{\it e}&{}&d&{}&b&{}&1&{}&{\ddots}
 \\[-1.8mm]{{}}{}&{}&{}&{}&{}&b&{}&d&{}&d&{}&b&{}&1&{}&1&{}
 \\[-1.8mm]{{}}{}&{}&{}&{}&{}&{}&b&{}&c&{}&b&{}&1&{}&1&{}&{{\ddots}}
 \\[-1.8mm]{{}}{}&{}&{}&{}&{}&{}&{}&b&{}&b&{}&1&{}&1&{}&{{\ddots}}&{}
 \\[-1.8mm]{{}}{}&{}&{}&{}&{}&{}&{}&{}&a&{}&1&{}&1&{}&{{\ddots}}&{}&{}
 \\[-1.8mm]{{}}{}&{}&{}&{}&{}&{}&{}&{}&{}&1&{}&1&{}&{{\ddots}}&{}&{}&{}
 \\[-1.8mm]{{}}{}&{}&{}&{}&{}&{}&{}&{}&{}&{}&{\ddots}&{}&{{\ddots}}&{}&{}&{}&{}
 \end {array}
 \right].
 \een\looseness=1
Having the latter structure and illustration in mind we can
contemplate any $R=2k+1$ and verify, by direct calculation, that the
rightmost unknown in the last line of Table~\ref{pexp4} (i.e., the
value of the wedge element~$B_k$) emerges from equation~(\ref{quas}) as
def\/ined, recurrently, at any one of the eight pairs of the
subscripts $(m,n)=(-2,\pm 1)$, $(m,n)=(-1,\pm 2)$, $(m,n)=(1,\pm 2)$
or $(m,n)=(2,\pm 1)$. Similarly, the leftmost and last unknown of
Table~\ref{pexp4} (i.e., the value of the central matrix element
$z_{2k+1}$) becomes determined, in terms of the freshly
pre-determined~$B_k$, by the $(m,n)=(-1,0)$ item of
equation~(\ref{quas}). Due to the symmetry of our ansatz for $\Theta$ we
could have also used $(m,n)=(0,\pm 1)$ or $(m,n)=(0,-1)$, with the
same result of course.

\section[Nonequivalent isospectral Hermitian Hamiltonians $\mathfrak{h}$]{Nonequivalent isospectral Hermitian Hamiltonians $\boldsymbol{\mathfrak{h}}$}\label{dyson}

In principle, there exist many non-equivalent factorizations of a
given quasilocal metric with  $(2R-1)\geq 3$ diagonals. In our
illustrative example we may consider, in general, the superposition
 \begin{gather}
 \Theta=\Theta(\alpha_1, \alpha_2, \ldots,
 \alpha_R)=
 \alpha_1\Theta_1+ \alpha_2\Theta_2+ \cdots
 +\alpha_R\Theta_R
 \label{metrR}
 \end{gather}
and  factorize $\Theta=\Omega^\dagger\Omega$.

\subsection[$R=1$ and the diagonal Dyson mappings $\Omega$]{$\boldsymbol{R=1}$ and the diagonal Dyson mappings $\boldsymbol{\Omega}$}

\looseness=1
By far the most popular reconstruction of the Dyson operator
$\Omega$ from a given metric $\Theta=\Omega^\dagger\Omega$ is based
on an {\it ad hoc} assumption of Hermiticity
$\Omega=\Omega^\dagger=\sqrt{\Theta}$ (cf., e.g., \cite{Jones}). For
our present class of band-matrix models $\Theta_R$ such an
assumption does not lead to any problems when the metric is
diagonal, $R=1$. For illustration let us select $\Theta_1$ of
equation~(\ref{metr1}). Postulating the diagonality and positivity of the
related Dyson map $\Omega_1$ in its factorization
$\Theta_1=\Omega_1^\dagger\Omega_1$ we arrive at the following
unique result,
 \ben
 \Omega_1=
 \left[ \begin {array}{ccccc} {}\ddots&{}&
&{}&{}\\[-2mm]
{}{}&1&{}&&{}\\[-2mm]
{}{}&{}& \sqrt{(1+{g})/{(1-{g})}}&{}&
\\
{}{}&{}&{}&1&{}\\[-2mm]
{}{}&{}&{}&{}&\ddots
\end {array} \right]
 \een
knowledge of which enables us to evaluate the related matrix of the
Hamiltonian in the physical Hilbert space ${\cal H}^{\rm (P)}$,
 \be
 \mathfrak{h}= \left[ \begin {array}{ccccccc} &\ddots&{}&{}&{}&{}&{}\\{}\ddots&2
&-1&{}&{}&{}&{}\\{}{}&-1&2&-\sqrt{1-{g}^2}&{}&{}&{}\\{}{}&{}
&-\sqrt{1-{g}^2}&2&-\sqrt{1-{g}^2}&{}&{}\\{}&{}&{}&-\sqrt{1-{g}^2}&2&-1&{}
\\{}{}&{}&{}&{}&-1&2&\ddots\\{}{}&{}&{}&{}&{}&\ddots&\ddots
 \end {array} \right]\,.
 \label{diago}
 \ee
The of\/f-diagonal elements
$-\sqrt{1-{g}^2}=-1+(1-\sqrt{1-{g}^2})=-1+{g}^2/(1+\sqrt{1-{g}^2})$
are composed of the kinetic-energy term $(-1)$ and a positive function
of the coupling.

\subsection[$R=2$ and a tridiagonal Dyson mapping $\Omega$]{$\boldsymbol{R=2}$ and a tridiagonal Dyson mapping $\boldsymbol{\Omega}$}

{\sloppy
We should issue a warning that the usual assumption of Hermiticity
of factors $\Omega= \Omega^\dagger$ would be rather
counterproductive because the Hermitian square root of our sparse
matrices $\Theta$ would be a non-sparse matrix. The quasilocality
property (i.e., a convergence to locality during the limiting
transition $h\to 0$) would be lost for the Dyson map. In opposite
direction, when we preserve the quasilocality (i.e., the band-matrix
structure) of Dyson matrices, we achieve, at any nonzero $h>0$, a
signif\/icant simplif\/ication of the physical representation of our
operators of observables in~${\cal H}^{\rm (P)}$.

}

Without any signif\/icant loss of generality the merits of such a
strategy may be illustrated via the most general tridiagonal metric
$\Theta(\gamma)=2\Theta_1+\gamma\,\Theta_2$. First of all we have to
guarantee its positivity and invertibility but both these properties
are easily shown to be guaranteed inside the open interval of
$\gamma \in (-1,1)$.

Secondly, we have to demonstrate the {\em practical} feasibility of
the rather dif\/f\/icult extraction of a quasilocal factor $\Omega$ from
a given quasilocal $\Theta= \Omega^\dagger\Omega$. For this purpose,
let us slightly simplify the presentation of the argument and choose
$\gamma\to -1$ lying on the  boundary of the open domain of
existence of the metric operator (\ref{metrR}). In this limiting
case our doubly inf\/inite metric acquires a rather elementary matrix
form,
 \ben
 \Theta=\left[ \begin {array}{ccccccccc}
 {\ddots}&{\ddots}&{}&{}&{}&{}&{}&{}&{}\\
{}{\ddots}&2&-1&{}&{}&{}&{}&{}&{}\\
{}{}&-1&2&-1&{}&{}&{}&{}&{}\\
{}{}&{}&-1&2&-1-{g}&{}&{}&{}&{}\\
{}{}
   &{}&{}&-1-{g}&2\,{\frac{1+{g}}{1-{g}}}&-1-{g}&{}&{}&{}\\
{}{}&{}&{}&{}&-1-{g}&2&-1&{}&{}\\
{}{}&{}&{}&{}&{}&-1&2&-1&{}\\
{}{}&{}&{}&{}&{}&{}&-1&2&{\ddots}\\
{}{}&{}&{}&{}&{}&{}&{}&{\ddots}&{\ddots}
\end {array}
\right].
 \een
This matrix is most easily factorized into band-matrix Dyson-mapping
factors, $\Theta=\Omega^\dagger\Omega$ when we postulate the
following special form of the Dyson-mapping matrix which is
asymmetric and tridiagonal,
 \be
 \Omega=
 \left[ \begin {array}{cccc|c|cccc}
  {\ddots}&{\ddots}&&&&&&&\\
 {}{}&1&-1&&&&&&\\
 {}&&1&-1&&&&&\\
 {}&&&1&-1-{g}&&&&\\
 \hline
 {}&&
   &&\sqrt{\frac{2{g}^2(1+{g})}{1-{g}}}&&&&\\
   \hline
{}&&&&-1-{g}&1&&&\\
{}&&&&&-1&1&&\\
{}&&&&&&-1&1&{}\\
{}&&&&&&&{\ddots}&{\ddots}
 \end{array} \right].
 \label{exapp}
 \ee
During the transition to a measurable coordinate $y_m$ appearing as
the argument in the physical wave function $\psi^{\rm (P)}(y_m) \in
{\cal H}^{\rm (P)}$, the similar quasilocal Dyson mappings enter the
scene via f\/inite sums
 \be
 \psi^{\rm (P)}(y_m) := \sum_{n}
  \Omega(y_m,x_{n})\psi^{\rm (F)}(x_{n}) .
  \label{hugone}
 \ee
In an ef\/fective theory as advocated in~\cite{Jonesdva}, both
the arguments $x_m$ and $y_m$ may be considered, in some sense,
observable. Then, equation~(\ref{hugone}) introduces just a certain
``smearing'' of coordinates, at the non-vanishing lattice sizes $h>0$
at least. The extent of this smearing is proportional to the number
of diagonals in $\Omega$.

We have to remember that we choose the parameter $\gamma=1$ which
lies, strictly speaking, {\em out of} the open interval of its
admissible values. Due care is needed when working with this option.
Nevertheless, a part of its undeniable methodical appeal still
recurs with the easiness of the evaluation of the inverse
 \ben
 \Omega^{-1}=
 \left[ \begin {array}{cccc|c|cccc}
  {\ddots}&{\ddots}&\ddots&\vdots&\vdots&&&&\\
 {}{}&1&1&1&\sqrt{\frac{1-{g}^2}{2{g}^2}}&&&&\\
 {}&&1&1&-\sqrt{\frac{1-{g}^2}{2{g}^2}}&&&&\\
 {}&&&1&\sqrt{\frac{1-{g}^2}{2{g}^2}}&&&&\\
 \hline
 {}&&
   &&\sqrt{\frac{1-{g}}{2{g}^2(1+{g})}}&&&&\\
   \hline
{}&&&&\sqrt{\frac{1-{g}^2}{2{g}^2}}&1&&&\\
{}&&&&-\sqrt{\frac{1-{g}^2}{2{g}^2}}&1&1&&\\
{}&&&&\sqrt{\frac{1-{g}^2}{2{g}^2}}&1&1&1&{}\\
{}&&&&\vdots&\vdots&\ddots&{\ddots}&{\ddots}
 \end{array} \right]\,.
 \een
The extreme $\gamma=1$ example also of\/fers a very useful guide for
transition to the generic  $|\gamma|<1$ where an ansatz for
$\Omega(\gamma)$ may be made with the same sparse-matrix structure.
This would lead to an ef\/f\/icient factorization  recipe based on the
use of continued fractions in the def\/inition of inverse matrix (at
$R=2$, cf.\ its sample in~\cite{cf}). A generalization of this
algorithm to $R>2$ using the so called extended continued fractions
also exists (cf.~\cite{ecf}).

Let us now return to our $R=2$ schematic example with the initial
Hamiltonian $H$ of equation~(\ref{slef}) and with the  tridiagonal metric
$\Theta(\gamma)$ where $\gamma=1$. Easily we may show, by direct
computation, that the related isospectral Hermitian Hamiltonian
$\mathfrak{h}$ reads
 \ben
 \mathfrak{h}=
 \left[ \begin {array}{ccc|ccc|ccc}
  {\ddots}&{\ddots}&{}&{}&{}&{}&{}&{}&{}
\\
{\ddots}&2&-1&{}&{}&{}&{}&{}&{}\\
{}&-1&2&-1&{}
&{}&{}&{}&{}\\
\hline
 {}&{}&-1&2-{{g}}^{2}&-\sqrt{2{g}^2(1\!-\!{g}^2)} &1-{{g}}^{2}&{}&{}&{}\\
 {}&{}&{}&-\sqrt{2{g}^2(1\!-\!{g}^2)} &2\,{{g}}^{2}&-\sqrt{2{g}^2(1\!-\!{g}^2)} &{}&{}&{}
\\
{}&{}&{}&1-{{g}}^{2}&-\sqrt{2{g}^2(1\!-\!{g}^2)} &2-
{{g}}^{2}&-1&{}&{}\\
\hline
 {}&{}&{}&{}&{}&-1&2&-1&{}
\\
{}&{}&{}&{}&{}&{}&-1&2&{\ddots}\\
{}&{}&{}&{}&{}&{} &{}&{\ddots}&{\ddots}\end {array} \right] .
 \een
We see that it remains sparse but dif\/ferent from its diagonal-metric
predecessor (\ref{diago}).

\section{Discussion} \label{summary}

Beyond the horizons given by our present sample of ${\cal PT}$-symmetric $H$ we may expect that many other models would also
prof\/it from the $h>0$ approximation treating the coordinate $x$ as a~discretized quantity $x_k=hk$. This approach enabled us to construct
the $(2R-1)$-diagonal metrics at all $R$ and it also allowed us to
postpone the study of the continuous coordinate limit $h\to 0$ to
the very end of all the calculations. Let us note now that this
limiting transition need not necessarily be easy. Sometimes, it may
be necessary to replace the simple-minded function $\Theta(x,x')$ of
two real variables  in formula (\ref{e777}) by a suitable (e.g.,
momentum-dependent) operator generalization.

At a f\/ixed level of approximation $h>0$ one can usually skip the
dif\/f\/icult discussions of the continuous coordinate limit. Even then,
due care must be paid to the (approximate) discrete theory where the
values $x_n$ of the individual coordinates play the role of
arguments in wave functions $\psi^{\rm (F)}(x_n)\in {\cal H}^{\rm (F)}$ {\em
without} probabilistic interpretation. The same discretized
coordinates also enter the correspondence-principle-ref\/lecting
def\/initions of the potential and of the kinetic-energy operator
$\triangle$ of equation~(\ref{dvanact}). All of these operators certainly
carry the decisive information about the dynamics of the quantum
system, provided only that the wave functions $\psi^{\rm (F)}(x_n)$ are
mapped, into the physical space ${\cal H}^{\rm (P)}$, according to the
following diagram,
 \ben
  \ba
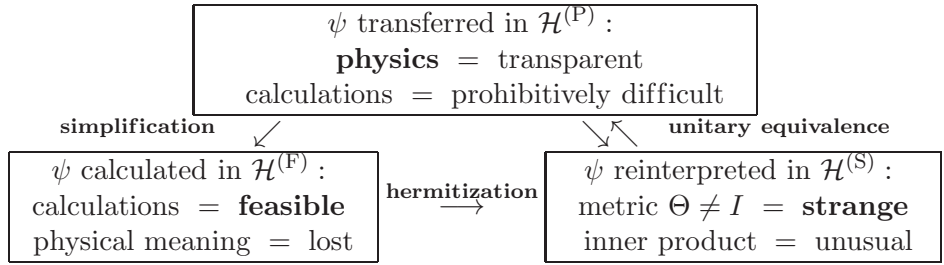
    \begin{array}{|c|}
 \hline
  \psi\
 {\rm transferred \ in\ } {\cal H}^{\rm (P)}:\
 \\
  \ \ \ \ {\rm  {\bf physics}\  = \ transparent } \ \\
  \ \ \ \ {\rm  calculations\ =\ prohibitively\ dif\/f\/icult\ } \ \\
 \hline
 \ea
 \\
 \stackrel{\bf  simplif\/ication}{}
 \ \ \ \
  \swarrow\ \  \  \ \ \ \ \ \
 \ \ \ \ \ \ \ \
  \ \  \  \ \ \ \ \ \
 \ \ \ \ \  \searrow \nwarrow\ \ \
 \stackrel{\bf  unitary\ equivalence}{}\\
 \begin{array}{|c|}
 \hline%
 \psi\
  {\rm calculated\ in\ } {\cal H}^{\rm (F)}:\ \\
 \
  {\rm  calculations\ = \ {\bf  feasible}\ }\    \\ %
 \
  {\rm  physical\ meaning \ = \ lost}\  \\
  \hline
 \ea
 \stackrel{ {\bf  hermitization}  }{ \longrightarrow }
 \begin{array}{|c|}
 \hline \psi\
  {\rm reinterpreted \ in\ } {\cal H}^{\rm (S)}:\
 \\
  \ \
  {\rm metric\ }
  \Theta\neq I\ =
   \  {\rm \bf strange\ } \  \\ %
  \ \
  {\rm   inner\ product\ = \ unusual}\   \\ %
 \hline
 \ea
\\
\\
\ea
 \een
Physical predictions necessitate a transfer of wave functions from
the computation-facilitating Hilbert space ${\cal H}^{\rm (F)}$ to the
correct physical Hilbert space ${\cal H}^{\rm (P)}$. The {\em ambiguity}
of this transfer is well illustrated by equation~(\ref{he-man})
containing the set of  free parameters $\lambda_j$ and $\kappa_j$
which ref\/lects their presence also in the related Dyson map
$\Omega=\Omega(\lambda_j,\kappa_j)$.

The main consequence of such a departure from the dictum of
textbooks is that for a quantized point particle our construction
and knowledge of wave functions must be complemented by a~suitable
upgrade of Hilbert space, i.e., usually, of the unphysical
${\mathbb L}^2({\mathbb R}) :={\cal H}^{\rm (F)}$. In our present models it has to
be accompanied by the limiting transition $h\to 0$. In this way the
unphysical version of the discrete normalization condition
 \be
 \sum_{k=0}^\infty  \psi^*(x_k) \psi(x_k)=1\qquad {\rm in}
 \quad {\cal H}^{\rm (F)},
 \label{jednac}
 \ee
will be replaced by its continuous limit (cf.\ equation~(\ref{diracs}))
while, similarly, its physical, ``standardized" counterpart
 \be
 \sum_{j=0}^\infty  \sum_{k=0}^\infty  \psi^*(x_j)
 \Theta(x_j,x_k) \psi(x_k)=1\qquad {\rm in}  \quad {\cal H}^{\rm (S)}
 \label{laots}
 \ee
will be assumed convergent to a double integral formula with $h\to
0$ (cf.\ equation~(\ref{e777})).

In this setting the main message of our present paper can be read as
a proposal of transition from the diagonal Runge--Kutta metrics (say,
of~\cite{FT,prd}) to their non-diagonal, $(2R-1)$-diagonal
sparse-matrix generalizations with $R\geq 2$. Under this
generalization the latter assumption of convergence
$\Theta(x_j,x_k')\to \Theta(x,x')$ need not always be satisf\/ied
since, in general, we have to expect the emergence of a more
complicated structure in the quasilocal metric $\Theta = \lim\limits_{h\to
0}\Theta(x_j,x_k')$. Typically, in a way inspired by the inspection
of equation~(\ref{he-man}) we might expect the emergence of a complicated
dependence of our quasilocal metrics on momenta, etc.

The study of the criteria distinguishing the metrics with a smooth
$x$-dependence (i.e., with operators represented by the mere
functions of two variables $\Theta(x,x')$) from  the other operators
of metrics without such an elementary representation lies far beyond
the scope of our present paper. Only a few comments may be added.

In the f\/irst one we repeat that the physical probabilistic
interpretation of the system requires the introduction of the
``standard'' Hilbert space ${\cal H}^{\rm (S)}\neq {\mathbb L}^2({\mathbb R})$
where the time-evolution becomes unitary. Although this space may be
def\/ined as spanned by the same functions of $x$, their norm must be
def\/ined dif\/ferently. This postulate degenerates back to the standard
textbook scenario when one returns to the Dirac's local metric,
$\Theta(x,x')\to \delta(x-x')$. {\it Vice versa}, for non-Hermitian
$H \neq H^\dagger$ with real spectra one may, sometimes, succeed in
constructing the corresponding {\it ad hoc} metric kernel
$\Theta(x,x')\neq \delta(x-x')$.

Our second comment will emphasize that our work with
dif\/ference-equation approximants has been motivated by the tedious
nature of some alternative perturbation-expansion techniques to
Schr\"{o}dinger equations, say, of~\cite{Jones,Jonesdva}. In
the conclusion let us mention a particularly interesting possibility
of a change of perspective. It emerged with the publication of paper
\cite{aliKG} where the f\/irst-quantized Klein--Gordon equation has
been considered. In this case, the three-Hilbert space formulation
of quantum mechanics of\/fers certain interesting new possibilities of
the interpretation of the role of individual spaces. The point is
that the initial Hamiltonian  $H\neq H^\dagger$ acquired the
physical meaning (i.e., relativistic covariance) in ${\cal H}^{\rm (F)}$
rather than in ${\cal H}^{\rm (P)}$. In our present language this would
lead to the  modif\/ied theoretical arrangement of our three Hilbert
spaces,
 \ben
  \ba
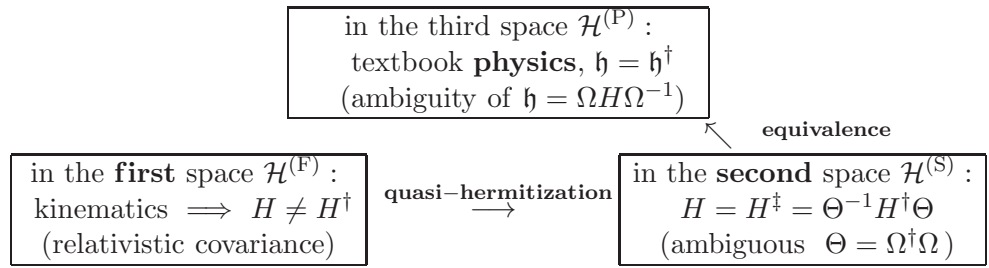
    \begin{array}{|c|}
 \hline
 {\rm in\ the\ third\ space\ } {\cal H}^{\rm (P)}:
 \\
  \ \ \ \ {\rm textbook  \ {\bf physics},\ } \mathfrak{h}=\mathfrak{h}^\dagger\ \\
  \ \ \ \ ({\rm ambiguity\ of\  } \mathfrak{h}=\Omega  H \Omega^{-1}) \ \\
 \hline
 \ea
\\
 \stackrel{\bf  }{}
\ \ \ \
  \ \  \  \ \ \ \ \ \
  \ \  \  \ \ \ \ \ \
  \ \  \  \ \ \ \ \ \
  \ \  \  \ \ \ \ \ \
  \ \  \  \ \ \ \ \ \
 \ \ \ \ \ \ \ \
\ \ \ \ \   \nwarrow\ \ \
 \stackrel{{\bf  equivalence} }{} \\
 \begin{array}{|c|}
 \hline%
  {\rm in\ the\  {\bf f\/irst\ } space\ } {\cal H}^{\rm (F)}:\ \\
 \
  {\rm {kinematics}}\ \Longrightarrow\  H \neq H^\dagger \\ %
 \
  {\rm  (relativistic\ covariance)}\  \\
  \hline
 \ea
 \stackrel{ {\bf\  quasi-hermitization }}{ \longrightarrow }
  \begin{array}{|c|}
 \hline%
  {\rm in\ the\ {\bf second\ }space\  } {\cal H}^{\rm (S)}:
 \\  \ H = H^\ddagger = \Theta^{-1}H^\dagger\Theta
    \\ %
  \ \
  {\rm  (ambiguous\ }\  \Theta=\Omega^\dagger \Omega\, {\rm ) }\   \\ %
 \hline
 \ea
\\
\ea
 \een
New ways towards old problems could be sought/found in this
direction. For example, in the light of some recently obtained new
results on the f\/irst quantization of relativistic particles with
spin~\cite{Smejkal}, an extension of these studies to a scattering
arrangement (e.g., along the lines indicated in our present paper)
would be a particularly challenging task.

\subsection*{Acknowledgements}

The author appreciates the support by the Institutional Research
Plan AV0Z10480505, by the M\v{S}MT ``Doppler Institute" project Nr.
LC06002 and by GA\v{C}R grant Nr. 202/07/1307.

\pdfbookmark[1]{References}{ref}
\LastPageEnding

\end{document}